\documentclass[useAMS,usenatbib,usegraphicx]{mn2e}

\title[Smooth matter and source size in microlensing simulations of gravitationally lensed quasars]{Smooth matter and source size in microlensing simulations of gravitationally lensed quasars}
\author[N. F. Bate, R. L. Webster and J. S. B. Wyithe]{N. F. Bate$^{1}$\thanks{E-mail:
nbate@physics.unimelb.edu.au}, R. L. Webster$^{1}$\thanks{E-mail: rwebster@physics.unimelb.edu.au} and J. S. B. Wyithe$^{1}$\thanks{E-mail: swyithe@physics.unimelb.edu.au}\\
$^{1}$School of Physics, The University of Melbourne, Parkville, Vic, 3010, Australia}
\begin{document}

\date{Accepted 2007 August 07. Received 2007 August 07; in original form 2007 January 29}

\pagerange{\pageref{firstpage}--\pageref{lastpage}} \pubyear{2007}

\maketitle

\label{firstpage}

\begin{abstract}
Several gravitationally lensed quasars are observed with anomalous magnifications in pairs of images that straddle a critical curve. Simple theoretical arguments suggest that the magnification of these images should be approximately equivalent, whereas one image is observed to be significantly demagnified. Microlensing provides a possible explanation for this discrepancy. There are two key parameters when modelling this effect. The first, the fraction of smooth matter in the lens at the image positions, has been explored by \citet{sw02}. They have shown that the anomalous flux ratio observed in the lensed quasar MG 0414+0534 is {\it a priori} a factor of 5 more likely if the assumed smooth matter content in the lens model is increased from 0\% to 93\%. The second parameter, the size of the emission region, is explored in this paper, and shown to be more significant. We find that the broadening of the magnification probability distributions due to smooth matter content is washed out for source sizes that are predicted by standard models for quasars. We apply our model to the anomalous lensed quasar MG 0414+0534, and find a 95\% upper limit of $2.62 \times 10^{16} h^{-1/2}_{70} (M/M_{\odot})^{1/2} cm$ on the radius of the I-band emission region. The smooth matter percentage in the lens is unconstrained.
\end{abstract}


\section{Introduction}
Gravitational microlensing, observed in some multiply-imaged quasars, offers an excellent opportunity to constrain the size and structure of the regions responsible for emitting the radiation. Some lensed quasars exhibit anomalous flux ratios between pairs of images straddling a caustic that cannot be convincingly explained using simple macrolens models. However microlensing models, particularly those that include a significant smooth matter component in the lens, offer a possible solution \citep*[][hereafter SW02]{sw02}.

The quadruply imaged quasar MG 0414+0534 is one example of a lensed quasar displaying anomalous flux ratios. These anomalies are observed in ratios between images $A_2$, located at a saddle point in the time delay surface, and $A_1$, located at a minimum. \citet{sm93} reported an $A_2/A_1$ I-band flux ratio of $0.45 \pm 0.06$ in observations taken on 2-4 November 1991. This result was supported by subsequent observations: $A_2/A_1 = 0.3 \pm 0.1$ from CFHT in the I-band on 1 March 1992 \citep{a94} and $A_2/A_1 = 0.47 \pm 0.01$ from HST in the I-band on 8 November 1994 \citep*{f97}. 

However, lensing theory tells us that image magnification scales as the inverse of perpendicular distance from a critical curve (\citealt{cr79}, \citealt{bn86}). Models that fit the observed image positions in MG 0414+0534 place images $A_1$ and $A_2$ either side of such a critical curve, and thus we would naively expect their magnification ratio, or flux ratio, to be $\sim1$ \citep*[][hereafter WMS95]{wms95}. Indeed an 8 GHz radio flux ratio of $A_2/A_1 = 0.90 \pm 0.02$ was observed on 2 April 1990 \citep{kh93}. This radio observation was much closer to the $A_2/A_1\sim1$ ratio predicted using lensing models (WMS95).

A mix of smooth and clumped matter distributions offers a potential explanation for the discrepancy between optical and radio flux ratios. Microlensing simulations that assume all matter in the lensing galaxy to be in compact objects yield a probability of 0.068 for a flux ratio lower than $A_2/A_1 = 0.45 \pm 0.06$ (WMS95). However, the addition of a smooth matter component in the lensing galaxy can significantly increase this probability up to values as large as 0.35, for sources with a characteristic radii much smaller than an Einstein Radius (SW02). The most compelling alternative explanation is millilensing by CDM substructure (\citealt{mm01}, \citealt{mz02}, \citealt{dk02}). However millilensing should also effect the radio emission, for which an anomalous flux ratio is not observed in MG 0414+0534.

Microlensing simulations have been used to place limits on the size of different emitting regions in several quasars. The majority of these efforts have focussed on the continuum region of Q2237+0305, and have relied on timescale arguments for high magnification events (eg, \citealt*{wps90}, \citealt{wfci91}, \citealt{wwtm00}). These arguments depend upon the apparent transverse velocity of the source, and stellar proper motions within the lens, neither of which are well known.

Upper limits on the size of emission regions can also be made using single observations of flux ratios between lensed images. These limits are free of the velocity dependency that arises in timescale arguments. For example WMS95 placed an 95\% upper limit of $3\times10^{16} (M/M_{\odot})^{1/2} cm$ on the I-band continuum emission region in MG 0414+0534 using this method. On larger scales, \citet*{wow05} placed an 80\% upper limit on the broad emission line region in Q2237+0305 of $2\times10^{17} (M/M_{\odot})^{1/2} cm$. In both cases these limits are large enough that the emission regions cannot be considered as point sources in microlensing simulations.

In this paper, we extend the investigation presented in SW02 to include the effects of varying source size on the magnification distributions of minimum and saddle point macroimages. We then use the anomalous observed flux ratio to place constraints on the size of the I-band emission region in MG 0414+0534, taking into account a varying smooth matter component in the lens.

Our simulation method is discussed in Section 2, followed by a qualitative examination of magnification histograms in Section 3. In Section 4 we use our microlensing simulations to place upper limits on the size of the I-band emission region in MG 0414+0534. We discuss the results of this investigation in Section 5.

Throughout this paper we use a cosmology with $H_0 = 70 km s^{-1} Mpc^{-1}$, $\Omega_m = 0.3$ and $\Omega_{\Lambda} = 0.7$.

\section{Simulations}
Microlensing simulations were conducted using a rayshooting method (eg, \citealt*{k86}, \citealt*{wpk90}). The key parameters for such simulations are the convergence $\kappa_{tot}$ and the shear $\gamma$ of the lens at the image locations. As discussed in SW02, the convergence of the lens can be split into two components -- a continuously distributed component $\kappa_c$, and a compact stellar component $\kappa_*$. We allowed the smooth matter percentage to vary from 0\% to 99\%.

\begin{table}
\caption{Lensing paramaters}
\label{lens_para}
\begin{tabular}{lcccl}
\hline
Image & Type & $\kappa_{tot}$ & $\gamma$ & $\mu_{tot}$ \\
\hline
M10 & minimum & 0.475 & 0.425 & 10.5 \\
S10 & saddle & 0.525 & 0.575 & -9.5 \\
$A_1$ & minimum & 0.472 & 0.488 & 24.2 \\
$A_2$ & saddle & 0.485 & 0.550 & -26.8 \\
\hline
\end{tabular}

\medskip
Lensing parameters for generic low magnification minimum and saddle point macroimages (M10 and S10 respectively, \citealt{sw02}), and for the images of interest in MG 0414+0534 ($A_1$ and $A_2$, \citealt{wms95}).
\end{table}

We conducted simulations for two sets of lensing parameters (see Table \ref{lens_para}). The first (labelled M10 and S10, following SW02) represent generic minimum and saddle point images with total magnifications $\mu\sim10$. These parameters were chosen both for their computational simplicity, and to allow comparison with SW02 results. The second set of parameters, taken from WMS95, describe the $A_1$ and $A_2$ images in MG 0414+0534.

Magnification maps were generated with a resolution of $2048 \times 2048$ pixels, covering an area of $24\eta_0 \times 24\eta_0$, where $\eta_0$ is the Einstein Radius projected on to the source plane ($3.75 \times 10^{16} h^{-1/2}_{70} (M/M_{\odot})^{1/2} cm$ for MG 0414+0534). Each pixel therefore has a side length of $0.01\eta_{0}$, which corresponds to $4.39 \times 10^{14} h^{-1/2}_{70} (M/M_{\odot})^{1/2} cm$ for MG 0414+0534.

The number of rays shot per unlensed pixel was chosen to give smooth probability distributions down to low magnifications. For the generic images 1000 rays per unlensed pixel were sufficient, whereas the higher magnification MG 0414+0534 images required 4000 rays per unlensed pixel. For each model, 20 magnification maps were generated per image. This number of $24\eta_0 \times 24\eta_0$ maps was found to provide enough statistically independent data points to construct reasonable probability distributions. 

The magnification maps were then convolved with a Gaussian source intensity profile, with characteristic radius varying from $0.05\eta_0$ (close to the source size used in SW02) to $2.00\eta_0$ in steps of $0.05\eta_0$. In physical units for MG 0414+0534, this corresponds to a characteristic source radius of $1.87 \times 10^{15} h^{-1/2}_{70} (M/M_{\odot})^{1/2} cm$ to $7.50 \times 10^{16} h^{-1/2}_{70} (M/M_{\odot})^{1/2} cm$.  Magnifications from the saddle point image were divided by those from the minimum image to construct flux ratios, or equivalently changes in magnitude using $\Delta m = 2.5log_{10}(\mu_2/\mu_1)$. Probability histograms were then derived from these changes in magnitude for each combination of smooth matter percentage and source radius.

\section{Flux Ratio Distributions}

\begin{figure}
  \includegraphics[width=80mm]{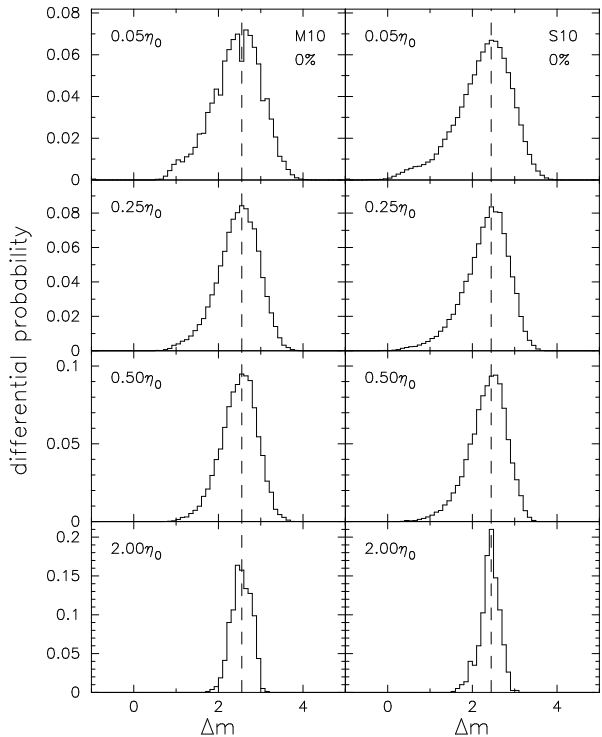}
  \caption{Differential probability histograms for $\Delta m$ for 0\% smooth matter percentage. Histograms are displayed for images M10 (left) and S10 (right), and four characteristic source radii (top to bottom). The dashed line indicates the predicated macro-model magnification in magnitudes.}
  \label{zero}
\end{figure}

\begin{figure}
  \includegraphics[width=80mm]{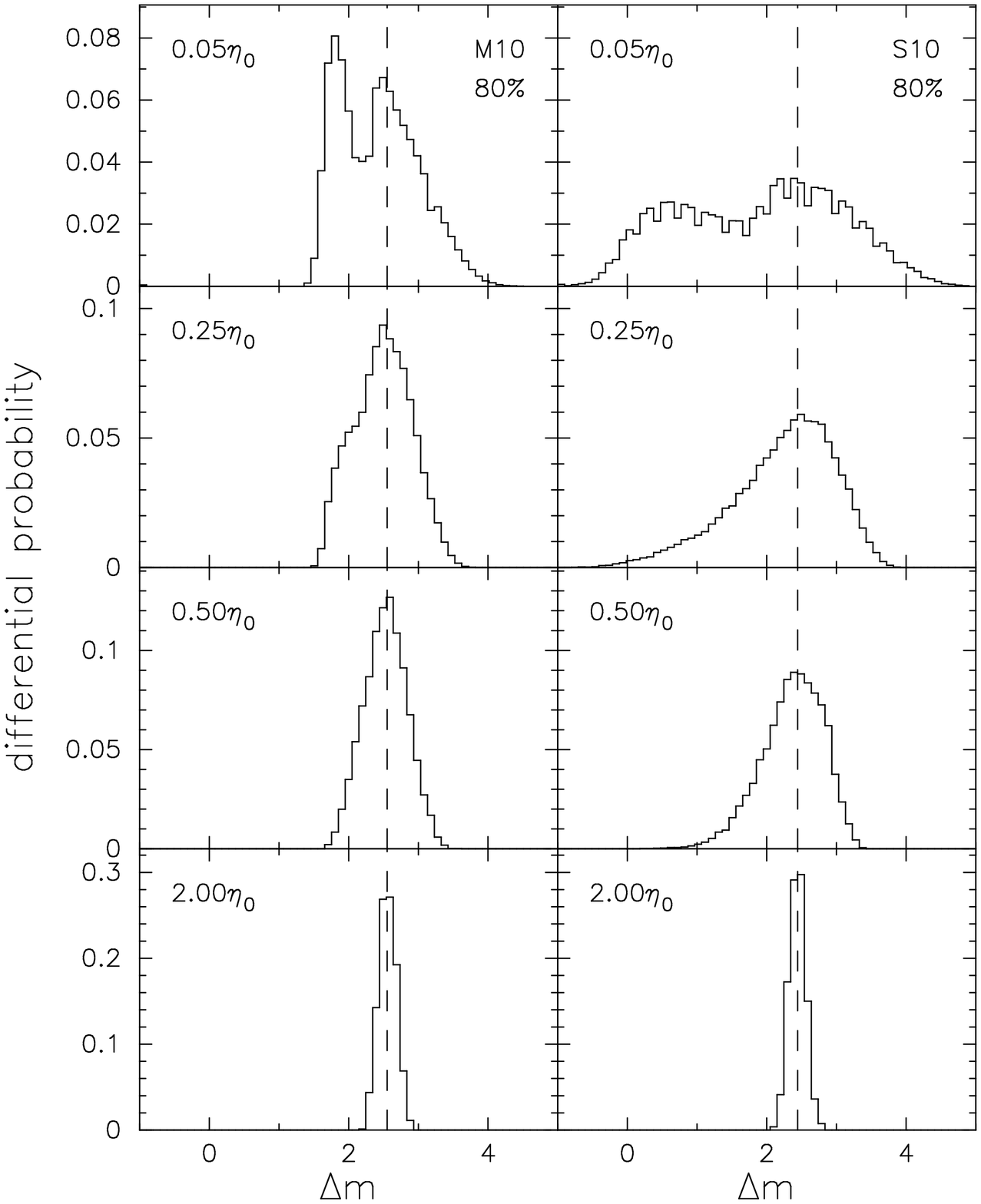}
  \caption{Differential probability histograms for $\Delta m$ for 80\% smooth matter percentage. Histograms are displayed for images M10 (left) and S10 (right), and four characteristic source radii (top to bottom). The dashed line indicates the predicated macro-model magnification in magnitudes.}
  \label{eighty}
\end{figure}

\begin{figure}
  \includegraphics[width=80mm]{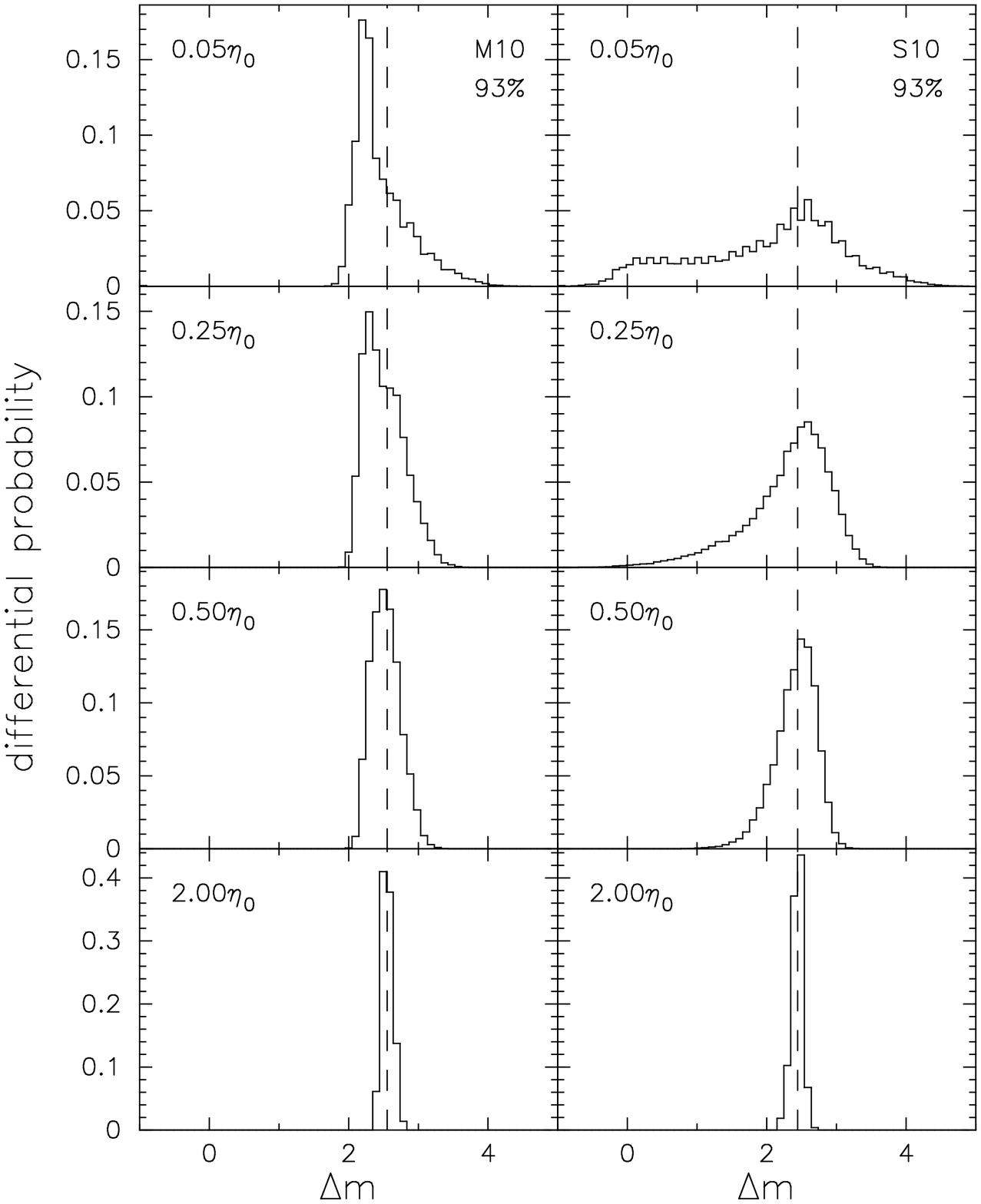}
  \caption{Differential probability histograms for $\Delta m$ for 93\% smooth matter percentage. Histograms are displayed for images M10 (left) and S10 (right), and four characteristic source radii (top to bottom). The dashed line indicates the predicated macro-model magnification in magnitudes.}
  \label{ninetythree}
\end{figure}

Fig. \ref{zero} shows differential probability distributions of change in magnitude $\Delta m$ for 0\% smooth matter content. Distributions are provided for the generic images M10 (minimum, left column) and S10 (saddle point, right column) and four characteristic source radii ($0.05\eta_0$, $0.25\eta_0$, $0.50\eta_0$ and $2.00\eta_0$).

For a source radius of $0.05\eta_0$, the minimum and saddle point distributions are essentially identical. As source radius is increased both distributions become narrower, while preserving their overall shape. This is expected; fluctuations in source magnification due to microlensing are strongly dependent on the size of the region being lensed, as has been noted many times in the literature (eg, \citealt{cr79}, \citealt{wps90}, \citealt*{wwt00a}, \citealt*{msw05}).

When smooth matter percentage is increased to 80\% (Fig. \ref{eighty}), the behaviour described in SW02 becomes apparent. For a characteristic source radius of $0.05\eta_0$ the minimum and saddle point histograms are quite different. Both distributions are bifurcated, with a second peak towards $\Delta m = 0$ (the unlensed case). Separate peaks correspond to different numbers of micro-images \citep*{gsw03}. In addition, the saddle point distribution covers a much broader range in $\Delta m$ than the minimum distribution.

Increasing the source radius in the 80\% smooth matter case has a more extreme effect on the shape of the probability distributions. For a characteristic radius of $0.25\eta_0$, the second peak in both distributions has disappeared. A long tail towards $\Delta m = 0$ remains in the saddle point distribution. This tail is suppressed by the time a characteristic source radius of $0.50\eta_0$ is reached. As in the 0\% case, increasing the source radius causes the distributions to narrow. However, at source radius $2.00\eta_0$ the 80\% smooth matter probability distributions are narrower than in the 0\% smooth matter case.

We find similar behaviour in the 93\% smooth matter case (Fig. \ref{ninetythree}). For a small source,  minimum and saddle point distributions appear quite different. As the source size is increased, the distributions quickly become narrow. The flat tail towards $\Delta m = 0$ in the $0.05\eta_0$ saddle point distribution is strongly suppressed by the time a source radius of $0.50\eta_0$ is reached. The combination of a high smooth matter component and a large source size make magnification fluctuations due to microlensing very minor.

These results differ from the predictions made in SW02. Although SW02 didn't present a detailed examination of the response of magnification histograms to increasing source size, they predict that doing so would compress the magnification histograms horizontally, while preserving their shapes. On the other hand, Fig. \ref{eighty} and Fig. \ref{ninetythree} show quite clearly that increasing source radius compresses the probability distributions, and washes out the broadening of the distribution for negative changes in magnitude. \citet*{cko07} observe the same behaviour for a different region of $\kappa-\gamma$ parameter space. As a result, a large smooth matter percentage can only explain anomalous flux ratios if the source size is small relative to the Einstein Radius.

\section{Limits on source size in MG 0414+0534}

\begin{figure*}
  \includegraphics[width=175mm]{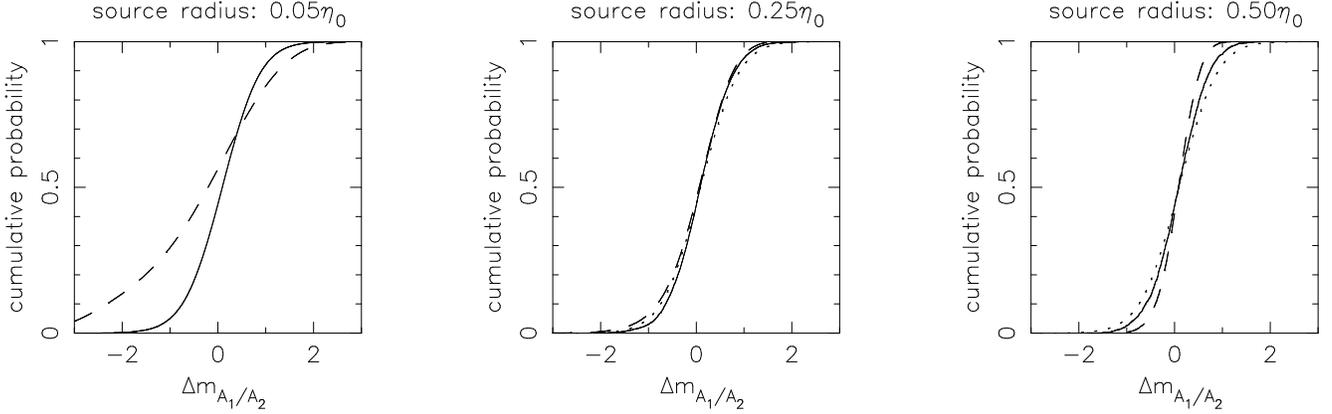}
  \caption{Cumulative probability distributions for $\Delta m$ between images $A_1$ and $A_2$ in MG 0414+0534. Smooth matter percentages of 0\% (solid line) and 93\% (dashed line) are displayed, for three characteristic source radii (left to right). The line representing the 0\% smooth matter, $0.05\eta_0$ source radius distribution is provided in the second and third panels (dotted line) for comparison.}
  \label{0414_hist}
\end{figure*}

We repeated the above analysis for microlensing parameters appropriate for MG 0414+0534 and observed similar results. Fig. \ref{0414_hist} illustrates the effect of smooth matter and source size on the probability of observing anomalously low flux ratios in MG 0414+0534. For a small source radius, the $\Delta m$ cumulative probability distribution is considerably broader in cases where the smooth matter percentage is high. Given a small source, a flux ratio as low as the observed ratio $(A_2/A_1)_{obs} = 0.45$ has a probability of 0.07 for the 0\% smooth matter case, and 0.31 for the 93\% case. At a source radius of $0.25\eta_0$, however, the 0\% and 93\% smooth matter distributions are virtually identical, and for larger radii the 93\% smooth matter distribution is found to be narrower than the 0\% smooth matter distribution.

We next compare the observed $A_2/A_1$ flux ratio for MG 0414+0534 with our model probability distributions and construct an a posteriori probability distribution for characteristic source radius and smooth matter percentage. 

Following WMS95 and SW02 we conduct our analysis using an observed flux ratio of $R_{obs} = (A_2/A_1)_{obs} = 0.45 \pm 0.06$ \citep{sm93}. By comparing this observed flux ratio with conditional probability distributions for the flux ratio, we constructed likelihoods for the observed ratio given varius source radii $\eta$, in units of the Einstein Radius $\eta_0$, and smooth matter percentages $s = \kappa_c/\kappa_{tot}$. Using Bayes' theorem, this likelihood $L(R_{obs}|s, \eta)$ was converted to an a posteriori differential probability distribution for smooth matter percentage and source radius as a fuction of $R_{obs}$.

\begin{equation}
\frac{d^2P}{dsd\eta}\Big |_{R_{obs}} \propto L(R_{obs}|s, \eta)\frac{dP_{prior}}{ds} \frac{dP_{prior}}{d\eta}
\label{bayes}
\end{equation}
These distributions are then marginalised over the observed distribution for flux ratio,

\begin{equation}
\frac{d^2P}{dsd\eta} = \int dR \left(\frac{d^2P}{dsd\eta}\Big |_{R_{obs}}\right) \frac{1}{\sqrt{2\pi}\Delta R_{obs}}\rmn{exp}(\alpha)
\label{bayes2}
\end{equation}

\begin{equation}
\rmn{with}\: \alpha = {\frac{-(R-R_{obs})^2}{2\Delta R_{obs}^2}}
\label{alpha}
\end{equation}
where the error in the flux ratio was treated as a Gaussian with characteristic radius equal to the observational error of $\pm 0.06$. We used a logarithmic Bayesian prior for source radius,  

\begin{eqnarray}
\nonumber
\frac{dP_{prior}}{d\eta} &\propto& \frac{1}{\eta} \hspace{5mm} \rmn{where} \hspace{5mm} \eta \leq 2.00\eta_0\\
&=& 0 \hspace{5mm} \rmn{otherwise}
\label{prior_eta}
\end{eqnarray}
and a constant Bayesian prior for smooth matter percentage,

\begin{eqnarray}
\nonumber
\frac{dP_{prior}}{ds} &\propto& 1 \hspace{5mm} \rmn{where} \hspace{5mm} 0 \leq s \leq 0.99 \\
&=& 0 \hspace{5mm} \rmn{otherwise}
\label{prior_kappa}
\end{eqnarray}

Source size is a quantity with units (in this case, units of Einstein Radius), and is therefore assumed to have a prior probability that is constant per unit logarithm. The choice of a prior that is flat in the logarithm ensures that the ratio of prior probability for two values of source size does not depend on the units chosen. On the other hand, the value of the smooth matter fraction is a dimensionless quantity and is therefore assumed to have a uniform prior.

\begin{figure}
  \includegraphics[width=84mm]{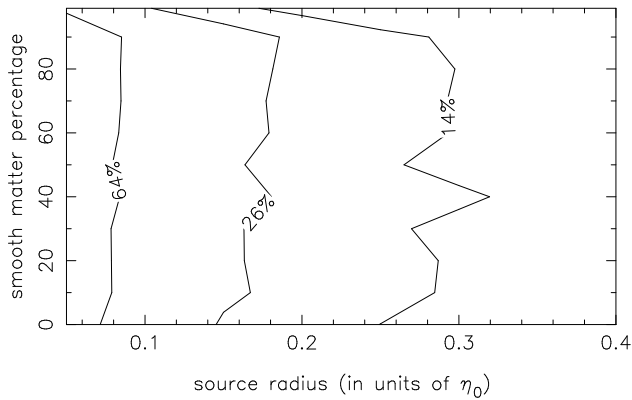}
  \caption{Probability distribution for characteristic source radius, in units of the Einstein Radius $\eta_0$, and smooth matter percentage. The contours show levels at 64\%, 26\% and 14\% of the peak likelihood. These contours were constructed for an observed flux ratio of $(A_2/A_1)_{obs} = 0.45\pm0.06$ and a logarithmic source radius prior.}
  \label{contour}
\end{figure}

Probability contours were then drawn through the resulting distribution and are plotted in Fig. \ref{contour}. This shows that our simulations do not constrain the smooth matter content of the lens. On the other hand an upper limit on the size of the I-band emission region exists for all smooth matter percentages.

We can therefore place a limit on the size of the I-band emission region in the quasar. To do so, we marginalise the differential probability distribution over smooth matter percentage: 

\begin{equation}
\frac{dP}{d\eta} = \int\frac{d^2P}{dsd\eta} ds
\label{overk}
\end{equation}
We then find the probability that the source is smaller than a particular radius, given the observed flux ratio:

\begin{equation}
P(<\eta) = \int\limits_0^{\eta}\frac{dP}{d\eta'}d\eta'
\label{overn}
\end{equation}

\begin{figure}
  \includegraphics[width=84mm]{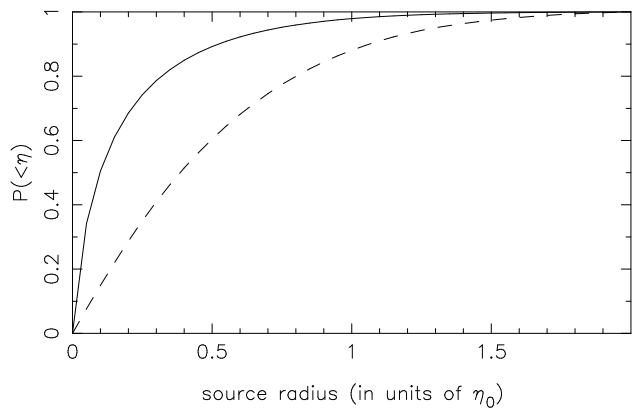}
  \caption{Cumulative probability that the I-band emission region in MG 0414+0534 is smaller than a radius $\eta$, given an observed flux ratio of $(A_2/A_1)_{obs} = 0.45\pm0.06$. Results are displayed for two source radius priors: logarithmic (solid line) and uniform (dashed line).}
  \label{probeta}
\end{figure}

Fig. \ref{probeta} shows the results of this analysis for an observed flux ratio of $(A_2/A_1)_{obs} = 0.45\pm0.06$ (solid line). We find that the radius of the I-band emission region in MG 0414+0534 is smaller than $0.70\eta_0$ with a statistical confidence of 95\%. In physical units, this limit is $2.62 \times 10^{16} h^{-1/2}_{70} (M/M_{\odot})^{1/2} cm$. Our limit on the I-band emission region is smaller than the limit found in WMS95. The improvement is due to the inclusion of a variable smooth matter component in our lens.

To test the sensitivity of our results to the assumed prior probability on source radius we have re-calculated the source radius constraints assuming a flat, rather than a logarithmic, prior. These results are presented in Fig. \ref{probeta} (dashed line). We find that the constraints are less stringent where the flat prior is used, with the data constraining the source radius to be smaller than about $1.3\eta_0$. Thus, while the data do constrain the upper limit on source radius, the precise constraints are sensitive to the prior chosen, indicating that the range of source radii that are consistent with the data remains considerable.

We have assumed a fixed microlens mass of $1.0M_{\odot}$ throughout this analysis. Previous studies have found the mean microlens mass to be a good approximation to the full microlens mass fuction (eg, \citealt*{wkr93}, \citealt{li95}, \citealt{wt01}). Under this assumption, a Salpeter mass function with a mass range of $0.1 < M/M_{\odot} < 10$ will reduce the physical size of our limit by a factor of $\sim2$, or more if the lower mass bound is decreased.

However, recent studies have shown that the interaction between source size and microlens mass function is complex. \citet{cko07} found that including a mass spectrum for the microlenses leads to broader probability distributions for larger source sizes. The range of the mass function is important; its slope is not. This will lead to a relaxation of our upper limit on source radius. \citet{lg06} demonstrated that small mass microlenses can mimic a smooth matter component if the source size is large enough. 

We have used a Gaussian source intensity profile throughout our analysis. This is reasonable as \citet{msw05} have shown that microlensing fluctuations are insensitive to all properties of the source model except the radius. We note, however, that in a sheared magnification map source ellipticity and position angle are important \citep{cko07}.

\section{Conclusion}
Anomalous flux ratios between close pairs of images have been observed in a number of lensed quasars. SW02 showed that including a large smooth matter percentage in lens models can significantly increase the probability of these anomalous flux ratios occurring, if the source radius is small. We confirm this result, but also find that increasing the source radius destroys the broadening of the magnification distributions in lenses with a smooth matter component. This result is also obtained by \citet{cko07}, for a different region of $\kappa-\gamma$ parameter space.

Using an observed I-band flux ratio of $(A_2/A_1)_{obs} = 0.45 \pm 0.06$ for the lensed quasar MG 0414+0534 \citep{sm93}, we place an upper limit (95\%) on the size of the emission region of $2.62 \times 10^{16} h^{-1/2}_{70} (M/M_{\odot})^{1/2} cm$. The smooth matter percentage in the lens model was allowed to vary between 0\% and 99\%, and remains unconstrained. This limit assumes a constant mass for the microlenses, and appears to be dominated by the source radius prior rather than the data.

Our size limit for MG 0414+0534 is consistent with limits on quasar continuum emission regions determined using other methods. For example, in the lensed quasar Q2237+0305, \citet{wwtm00} placed upper and lower limits (99\%) on the R-band emission region of $6 \times 10^{15}cm$ and $2 \times 10^{13}cm$ respectively. \citet{wps90} placed an upper limit on the Q2237+0305 optical emission region of approximately $2 \times 10^{15}cm$.

MG 0414+0534 is not the only lensed quasar displaying anomalous flux ratios in close images. PG 1115+080, SDSS J0924+0219, WFI J2026-4536, and HS 0810+2554 are four other well known examples \citep{p06}. An analysis similar to the one conducted here was undertaken for the broad emission line flux ratios in SDSS J0924+0219 using HST data \citep{k06}. The authors concluded that acceptable models exist with the broad line region as large as $\sim2.3 \times 10^{16} (M/M_{\odot})^{1/2} cm$ for a smooth matter percentage of 80-85\%. \citet{p06} found X-ray flux ratio anomalies in PG 1115+080 to be a factor of 6 more extreme than their optical counterparts, and concluded that the optical continuum emission region is ~10-100 times larger than expected from a thin accretion disk model.

By extending observations of MG 0414+0534 across a number of filters, we would be able to conduct our analysis for a series of different emission regions. Theories of quasar accretion suggest that different wavelengths are emitted from different regions in the source. Therefore with multi-band observations we could separate these regions in radius and directly probe the structure of the quasar central engine.

In conclusion, we find that differential magnification in lensed quasar images shows that microlensing is important, but that it does not measure the smooth matter content of the lens. The size of the emission region and parity of the images are more significant factors than the smooth matter component. We place a 95\% upper limit on the size of the I-band emission region in MG 0414+0534 of $2.62 \times 10^{16} h^{-1/2}_{70} (M/M_{\odot})^{1/2} cm$. Unlike most previous analyses, this limit is independent of the unknown transverse velocity of the source.

\section{Acknowledgements}
We would particularly like to thank Joachim Wambsganss for the use of his rayshooting code. We also thank the referee for insightful comments which substantially improved the presentation of these results. NFB acknowledges the support of an Australian Postgraduate Award.

\bsp

\label{lastpage}

\end{document}